\documentclass{article}

\usepackage{microtype}
\usepackage{graphicx}
\usepackage{subfigure}
\usepackage{amsmath}
\usepackage{mathtools}
\usepackage{amsfonts}
\usepackage{booktabs} 
\usepackage{lipsum}

\newcommand\blfootnote[1]{%
  \begingroup
  \renewcommand\thefootnote{}\footnote{#1}%
  \addtocounter{footnote}{-1}%
  \endgroup
}

\usepackage{hyperref}

\usepackage[accepted]{icml2019}

\icmltitlerunning{Unified framework for multivariate distributions in biological sequences}

\begin{document}

\twocolumn[
\icmltitle{Unified framework for modeling multivariate distributions in biological sequences}




\begin{icmlauthorlist}
\icmlauthor{Justas Dauparas}{fas}
\icmlauthor{Haobo Wang}{fas}
\icmlauthor{Avi Swartz }{college}
\icmlauthor{Peter Koo}{hhmi}
\icmlauthor{Mor Nitzan}{jhdsf}
\icmlauthor{Sergey Ovchinnikov}{jhdsf}
\end{icmlauthorlist}

\icmlaffiliation{fas}{FAS Division of Science, Harvard University}
\icmlaffiliation{jhdsf}{JHDSF Program, Harvard University}
\icmlaffiliation{hhmi}{Howard Hughes Medical Institute, Harvard University}
\icmlaffiliation{college}{Harvard College}


\icmlcorrespondingauthor{Mor Nitzan}{mornitzan@fas.harvard.edu}
\icmlcorrespondingauthor{Sergey Ovchinnikov}{so@fas.harvard.edu}

\icmlkeywords{Machine Learning, ICML}

\vskip 0.3in
]



\printAffiliationsAndNotice{} 


\begin{abstract}
Revealing the functional sites of biological sequences, such as evolutionary conserved, structurally interacting or co-evolving protein sites, is a fundamental, and yet challenging task. 
Different frameworks and models were developed to approach this challenge, including Position-Specific Scoring Matrices, Markov Random Fields, Multivariate Gaussian models and most recently Autoencoders. 
Each of these methods has certain advantages, and while they have generated a set of insights for better biological predictions, these have been restricted to the corresponding methods and were difficult to translate to the complementary domains.
Here we propose a unified framework for the above-mentioned models, that allows for interpretable transformations between the different methods and naturally incorporates the advantages and insight gained individually in the different communities. 
We show how, by using the unified framework, we are able to achieve state-of-the-art performance for protein structure prediction, while enhancing interpretability of the prediction process.

\end{abstract}

\section{Introduction}

The function of the products of biological sequences, such as proteins and RNAs, is, to a large extent, encoded directly into the sequence itself.  
However, determining the function from a single sequence remains a challenge. One promising approach is to leverage the power of evolution and comparison, to gain statistical power to identify which positions are conserved and which pairs of positions are coevolving, thus supporting their functional importance. 

Generative models are typically employed to infer such statistical properties from homologous sequences which are organized into a multiple sequence alignment (MSA). Such methods include Position-Specific Scoring Matrices (PSSMs), Markov Random Fields (MRFs), Multivariate-Gaussians (MGs), and more recently,  Autoencoders (AEs).  PSSMs are site-independent models that are limited to capture one-body terms, \textit{i.e.} evolutionary conservation, across the species in an MSA \cite{stormo1982use}. To capture additional two-body interaction terms, \textit{i.e.} coevolving positions, MRFs and MGs have emerged, specifically in the context of protein structure contact prediction \cite{balakrishnan2011learning, kamisetty2013assessing, ekeberg2013improved, morcos2011direct, baldassi2014fast}. MGs assume the distributions of the amino acids are continuous, and are able to capture higher-order interactions with their covariance matrix. On the other hand, MRFs explicitly model the amino acids as categorical distributions and capture higher order interactions with two-body coupling terms. 

Although MRFs and MGs have largely been considered to be two distinct approaches to model conservation and coevolution in protein sequences, we present a unified mathematical framework for both, formulated as a general autoencoder, reducing the differences between the two traditional approaches to trivial changes. Moreover, this unifying framework allows for the insights gained by one approach to be utilized and interpreted by the other.
We envision that this framework will enable merging the advantages of the different existing approaches, thus leading to better results for prediction of physical contacts and coevolution for biological sequences.  

\section{Results}
\begin{figure}
  \centering
  \includegraphics[height=3.7cm]{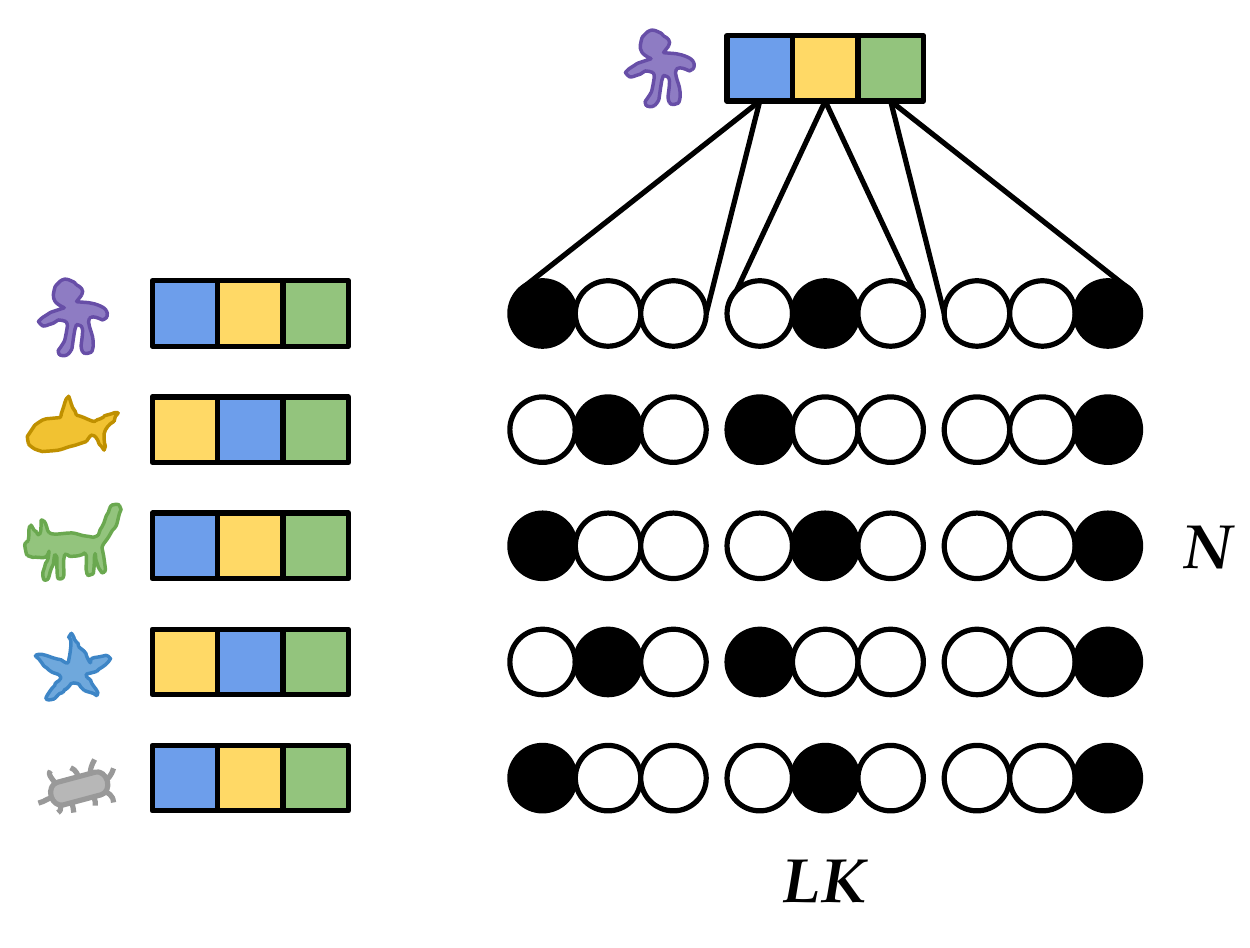}
  \caption{
  Aligned biological sequences such as DNA, RNA, or protein. Every position in the sequence is represented using one- hot encoding.
  The data matrix $\mathbf{X}\in \mathbb{R}^{N\times LK} = \mathbb{R}^{N\times L\times K}$ has $N$ sequences, every sequence is of length $L$ and every position in the sequence can have $K$ different states. 
  }
  \label{fig:1}
\end{figure}

For a set of homologous sequences in an MSA, $\mathbf{X}\in \mathbb{R}^{N\times L\times K}$, with $N$ sequences of length $L$ and an alphabet of size $K$, we would like to interpret the underlying patterns of those sequences; Which positions are conserved? Which subsets of positions are coevolving? The sequences can be encoded as a one-hot representation (Fig. \ref{fig:1}), where the sum over alphabet is equal to one, i.e. $\sum_{k=1}^K X_{nlk} = 1 \: \forall~{l,k}$.   

\subsection{Position-specific scoring matrix}\label{sec:pssm}

To capture site-independent patterns in biological sequences, \textit{i.e.} evolutionary conservation, a position-specific scoring matrix (PSSM) is frequently used \cite{stormo1982use}. The parameter in this model is a matrix $\mathbf{b}\in \mathbb{R}^{L\times K}$, and the loss function $\mathcal{L}_{PSSM}$ for this model can be written as an average cross entropy loss between the data and the PSSM matrix:
\begin{align}
    \mathcal{L}_{PSSM} &= \frac{1}{N}\sum_{n=1}^{N}\sum_{l=1}^{L}\operatorname{cce}[\mathbf{X},\operatorname{softmax}(\mathbf{b})],\\
    \mathcal{L}_{PSSM} &= -\frac{1}{N}\sum_{n=1}^{N} \sum_{l=1}^{L} \sum_{k=1}^{K} X_{nlk}\log{\frac{e^{b_{lk}}}{\sum_{q=1}^{K}{e^{b_{lq}}}}}
\end{align}
where the softmax and cross entropy functions are taken along the alphabet. See Fig. \ref{fig:2} for the graphical representation of the model. The maximum likelihood estimator for $\mathbf{b}$ is given by $b_{rs} = \log{\left(\frac{1}{N}\sum_{n=1}^{N}X_{nrs}\right)}$, \textit{i.e.} the natural logarithm of the empirical frequencies.

\subsection{MRF with pseudolikelihood is identical to cross entropy} \label{sec:mrf}

PSSMs only consider one-body terms, \textit{i.e.} conservations. A Markov Random Field, or Potts model, generalizes the PSSMs to also include two-body interaction terms \cite{lapedes1999correlated, thomas2005graphical, balakrishnan2009structure, weigt2009identification}. The maximum entropy solution to a model that is constrained to capture single site amino acid frequencies and pairwise amino acid frequencies in an MSA is a Boltzmann distribution given by $p(\mathbf{x}; \mathbf{W}, \mathbf{b}) = \frac{\tilde{p}(\mathbf{x}; \mathbf{W}, \mathbf{b})}{Z(\mathbf{W}, \mathbf{b})}$, where $\tilde{p}(\mathbf{x}; \mathbf{W}, \mathbf{b})$ is the unnormalized probability distribution and $Z(\mathbf{W}, \mathbf{b})=\sum_{\mathbf{x}} \tilde{p}(\mathbf{x};\mathbf{W}, \mathbf{b})$ is a normalization constant known as the partition function.  This model has parameters $\mathbf{b}\in \mathbb{R}^{L\times K}$ and $\mathbf{W}\in \mathbb{R}^{L\times K\times L\times K}$ where the reshaped $\mathbf{W}\in \mathbb{R}^{LK\times LK}$ matrix has zero values along the diagonal and is symmetric. To ensure zeros along the diagonal one can write $\left(\mathbf{1}_{LK\times LK}-\mathbf{I}\right)\circ\mathbf{W}$, where $\mathbf{1}_{LK\times LK}$ is the matrix of ones, $\mathbf{I}$ is an identity matrix, and $\circ$ denotes the Hadamard product.

The partition function for discrete distributions is given by a sum over all realizations of the data and hence it is often not tractable. However, the partition function cancels out for conditional probabilities, as achieved by the pseudo-likelihood objective \cite{besag1975statistical}:
\begin{align}
   \log{p(\mathbf{x})} &
   \approx\sum_{i=1}^n \log{\frac{\tilde{p}(x_i, \mathbf{x}_{-i})}{\sum_{x_i}\tilde{p}(x_i, \mathbf{x}_{-i})}},
\end{align}
where $\mathbf{x}$ has $n$ features, $x_i$ denotes the i-th feature and  $\mathbf{x}_{-i}$ all the features apart from the i-th one. 
A pseudo-likelihood approximation replaces the sum over all possibilities in the partition function only by the sum over the states of the considered location while all the other locations of the sequence are fixed from the data \cite{balakrishnan2011learning, kamisetty2013assessing, ekeberg2013improved}.

\begin{figure}
  \centering
  \includegraphics[height=3.5cm]{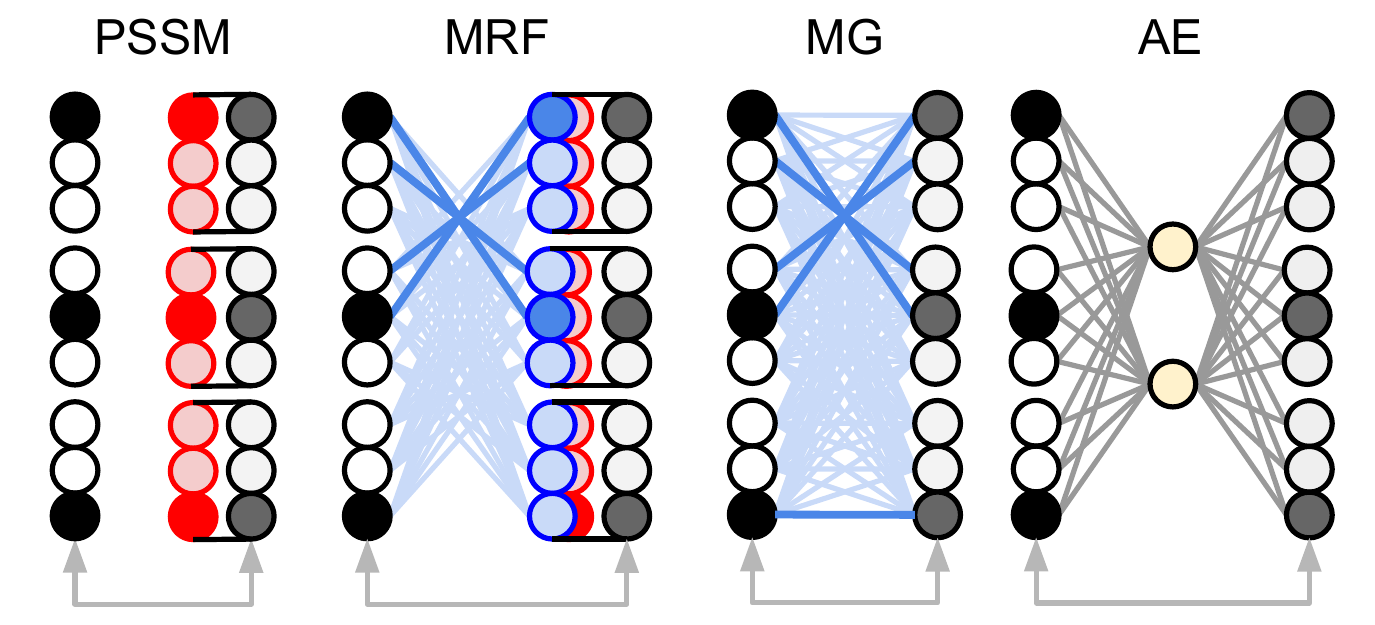}
  \caption{Graphical representation of the models. A position-specific scoring matrix (PSSM) model with the loss $\mathcal{L}_{PSSM}= \operatorname{mean}\left\{\operatorname{cce}[\mathbf{X},\operatorname{softmax}(\mathbf{b})]\right \}$ learns the parameters $\mathbf{b}\in \mathbb{R}^{L\times K} = \mathbb{R}^{LK}$ which are shown as a red vector. A Markov random field (MRF) with pseudo-likelihood has the loss $\mathcal{\hat{L}}_{MRF}= \operatorname{mean}\left\{\operatorname{cce}[\mathbf{X},\operatorname{softmax}(\mathbf{B}+\mathbf{X}\mathbf{W})]\right\}+\lambda \vert\vert\mathbf{W}\vert\vert_2^2$ where the parameters are $\mathbf{B}\in \mathbb{R}^{N\times LK}$ (red) and  $\mathbf{W}\in \mathbb{R}^{LK\times LK}$ (blue). A multivariate Gaussian (MG) type model has the loss $\mathcal{\hat{L}}_{MG} =\frac{1}{N} \vert\vert\mathbf{\hat{X}}-\mathbf{\hat{X}}\mathbf{W}\vert\vert_2^2+\lambda \vert\vert\mathbf{W}\vert\vert_2^2-2\gamma\operatorname{tr}(\mathbf{W})$ where $\mathbf{\hat{X}}\in \mathbb{R}^{N\times LK}$ is the mean-centered data and $\mathbf{W} \in \mathbb{R}^{LK\times LK}$ (blue) are the parameters. Either MRF or MG models can be extended to a full auto-encoder (AE) by addition of hidden layer(s).}
  \label{fig:2}
\end{figure}

The loss $\mathcal{L}_{MRF}$ for the MRF with a pseudo-likelihood approximation is given by
\begin{align}
\label{eq1}
\mathcal{L}_{MRF} = -\frac{1}{N}\sum_{n=1}^N\sum_{l=1}^L \log{\frac{e^{\sum_{k=1}^K X_{nlk} H_{nlk}}}{\sum_{q=1}^K e^{\sum_{k=1}^K C_{nqk} H_{nlk}}}}, 
\end{align}
where $H_{nlk}= B_{nlk}+\sum_{r=1}^{L}\sum_{s=1}^{K} X_{nrs} W_{rslk}$, the constant $\mathbf{C} \in \mathbb{R}^{N\times K\times K}$ is an encoding of alphabet which is broadcasted $C_{nqk}=c_{qk}, \: \forall~{n,l,k}$, and  $\mathbf{B}\in\mathbb{R}^{N\times L\times K}$ is a broadcasted $\mathbf{b}$, i.e. $ B_{nlk} = b_{lk} \: \forall~{n,l,k}$. 
If one-hot encoding is used then $\mathbf{c}$ is an identity matrix. The graphical representation of the model is shown in Fig. \ref{fig:2}.

We will now show that the loss function of an average cross entropy is identical to the Markov Random Field with one-hot encoding. The loss function $\mathcal{L}$ for the model using an average cross entropy loss is given by
\begin{align}
\label{eq2}
    \nonumber
    \mathcal{L} &= \frac{1}{N}\sum_{n=1}^{N}\sum_{l=1}^{L}\left\{\operatorname{cce}[\mathbf{X},\operatorname{softmax}(\mathbf{H})]\right\}_{nl},\\ \nonumber
   &=-\frac{1}{N}\sum_{n=1}^N\sum_{l=1}^L \log{ \frac{e^{\sum_{k=1}^{K}X_{nlk}H_{nlk}}}{\left(\sum_{g=1}^{K}e^{H_{nlg}}\right)^{\sum_{k=1}^{K} X_{nlk}}}},\\ 
   &=-\frac{1}{N}\sum_{n=1}^N\sum_{l=1}^L \log{ \frac{e^{\sum_{k=1}^{K}X_{nlk}H_{nlk}}}{\sum_{g=1}^{K}e^{H_{nlg}}}},
\end{align}
where the last simplification is a result of $\sum_{k=1}^K X_{nlk} = 1.$ Thus, Eq. \ref{eq2} is equal to Eq. \ref{eq1}, \textit{i.e.} $\mathcal{L}=\mathcal{L}_{MRF}$, which demonstrates that the MRF with a pseudolikelihood approximation is equivalent to the cce loss formulation.

\subsection{MG can be reformulated as mean squared error}

Suppose that each sequence is instead sampled from a multivariate Gaussian distribution, i.e. $\mathbf{x}^{(n)} \sim \mathcal{N}(\mathbf{0}, \mathbf{\Sigma})$, where $\mathbf{x}^{(n)}$ is the $n$th sequence in $\mathbf{X}$. 
The log-likelihood for i.i.d. $\mathbf{x}^{(i)}$ up to a constant is given by
\begin{align}
    \label{eq5}
    l(\mathbf{\Theta})&= \frac{1}{2} \log \operatorname{det}(\mathbf{\Theta})-\frac{1}{2 N} \sum_{n=1}^{N} \mathbf{x}^{(n) \top} \mathbf{\Theta} \mathbf{x}^{(n)},
\end{align}
where $\mathbf{\Theta}$ is the positive definite precision matrix defined as $\mathbf{\Theta} =\mathbf{\Sigma}^{-1}$. This is maximized when $\mathbf{\Theta}=\mathbf{S}^{-1}$,
where $\mathbf{S}$ is the empirical covariance matrix defined as $\mathbf{S} \coloneqq \frac{\mathbf{X}^{T}\mathbf{X}}{N}$.

We will now consider the  mean squared error loss function  of the centered data and the matrix multiplication of the centered data. This model can be thought of as a mapping onto itself with an affine transformation (see Fig. \ref{fig:2}). However, to prevent a trivial self-mapping, a regularization is needed.  

The model has parameters $\mathbf{W}\in \mathbb{R}^{LK\times LK}$ and the mean-centered data is denoted as $\mathbf{\hat{X}}=\mathbf{X}-\mathbf{\overline{X}}$, where  $\mathbf{\overline{X}}\in \mathbb{R}^{N\times LK}$ and $\overline{X}_{nf} = \frac{1}{N}\sum_{i=1}^{N}\overline{X}_{if} \: \forall{n}$. The loss function $\mathcal{L}_{MG}$ is defined as
\begin{align}
    \label{eq11}
    \nonumber
    \mathcal{L}_{MG} &=
    \frac{1}{N}\sum_{n=1}^{N}\sum_{f=1}^{LK}\{\operatorname{mse}[\mathbf{\hat{X}},\mathbf{\hat{X}}\mathbf{W}]\}_{nf}\\
    &=\frac{1}{N}\sum_{n=1}^{N}\sum_{f=1}^{LK}\left(\hat{X}_{nf}-\sum_{q=1}^{LK}\hat{X}_{nq}W_{qf}\right)^2.
\end{align}
To prevent the trivial identity solution, we add regularization $\mathcal{R}$ and write the total loss as $\mathcal{\hat{L}}_{MG} = \mathcal{L}_{MG} + \mathcal{R}$ where
\begin{align}
    \mathcal{R} &= \lambda \left\lVert  \mathbf{W}\right\rVert_2^2-2\gamma \operatorname{tr}(\mathbf{W}).
\end{align}
In this case the derivative of the total loss is
\begin{align}
    \frac{\partial\hat{\mathcal{L}}_{MG}}{\partial \mathbf{W}} 
    &=\frac{2}{N}\left(\mathbf{\hat{X}}^{T}\mathbf{\hat{X}}\mathbf{W}-\mathbf{\hat{X}}^{T}\mathbf{\hat{X}}\right)+2\lambda \mathbf{W}-2\gamma\mathbf{I}.
\end{align}
Solving $\frac{\partial\hat{\mathcal{L}}_{MG}}{\partial \mathbf{W}} =0$ leads to 
\begin{align}
    \mathbf{W}&=(\gamma-\lambda)\left(\operatorname{cov}(\mathbf{\hat{X}})+\lambda \mathbf{I}\right)^{-1}+\mathbf{I}
\end{align}
with the empirical covariance defined as $\operatorname{cov}(\mathbf{\hat{X}})\coloneqq\frac{\mathbf{\hat{X}^T}\mathbf{\hat{X}}}{N}$.
If we remove the identity from the parameter matrix by writing $\mathbf{W}=\mathbf{I}+\mathbf{\tilde{W}}$ then the modified loss up to a constant is given by
\begin{align}
    \label{eq3}
    \mathcal{\tilde{L}}_{MG}&= \frac{1}{N}\vert\vert\mathbf{\hat{X}}\mathbf{\tilde{W}}\vert\vert_2^2+\lambda \vert\vert\mathbf{\tilde{W}+\mathbf{I}}\vert\vert_2^2-2\gamma\operatorname{tr}(\mathbf{\tilde{W}})
\end{align}
and the solution to this optimization problem is simply shifted by the identity matrix, i.e.
\begin{align}
    \mathbf{\tilde{W}}
    &=(\gamma-\lambda)\left(\operatorname{cov}(\mathbf{\hat{X}})+\lambda \mathbf{I}\right)^{-1}.
\end{align}
If we choose $\gamma=1$ then $\mathbf{\tilde{W}}$ is given as the inverse of the regularized empirical covariance matrix, and if we set $\lambda=0$ then it is just the inverse of the covariance matrix. Therefore, the log-likelihood loss for the multivariate Gaussian model (Eq. \ref{eq5}) and the mean squared error loss (Eq. \ref{eq3}) both have the same global minimum. Interestingly, unlike for the multivariate Gaussian model, the mean squared error model does not require a postitive definite constraint. This switch can drastically reduce the run time.

If we write $\mathbf{W}\circ(\mathbf{1}_{LK\times LK}-\mathbf{I})$ in the MG model (Eq. \ref{eq11}) then the solution is given by
\begin{align}
\label{eq:13}
   \mathbf{W}&=\mathbf{I}-\mathbf{C}^{-1}\frac{\mathbf{I}}{\mathbf{C}^{-1}\circ \mathbf{I}},\\ \nonumber \mathbf{C}&=\operatorname{cov}(\mathbf{\hat{X}})+\lambda\mathbf{I}.
\end{align}
Therefore, the symmetric part is $\mathbf{\hat{W}}=(\mathbf{W}+\mathbf{W}^{T})/2$.
\subsection{Unifying multivariate distributions in a single framework}

In summary, we have demonstrated that: 1) Maximizing the pseudo-likelihood of a MRF is identical to minimizing the categorical cross entropy (CCE); 2) Maximizing the likelihood of a multivariate Gaussian model has the same global optima as minimizing mean-squared-error (MSE), where L2 regularization is identical to shrinking the covariance matrix before inversion. The significance of this is that  both models can be reformulated as a single dense (or fully-connected) layer that maps the input data onto itself. The difference is that MRF explicitly removes connections with the same positions, while MG includes these connections but must include regularization to prevent the trivial mapping. MRF and MG treat input data as categorical and continuous, respectively. Nevertheless, the benefit of reframing these problems as a graphical model enables  to leverage widely-used neural network frameworks, such as Tensorflow and PyTorch, which gives access to GPU libraries. Moreover, additional hidden layers can be added in between the inputs and outputs to enable auto-encoding (see Fig. \ref{fig:2}). 

\section{Application: Protein contact prediction}
\begin{figure}
  \centering
  \includegraphics[height=3.3cm]{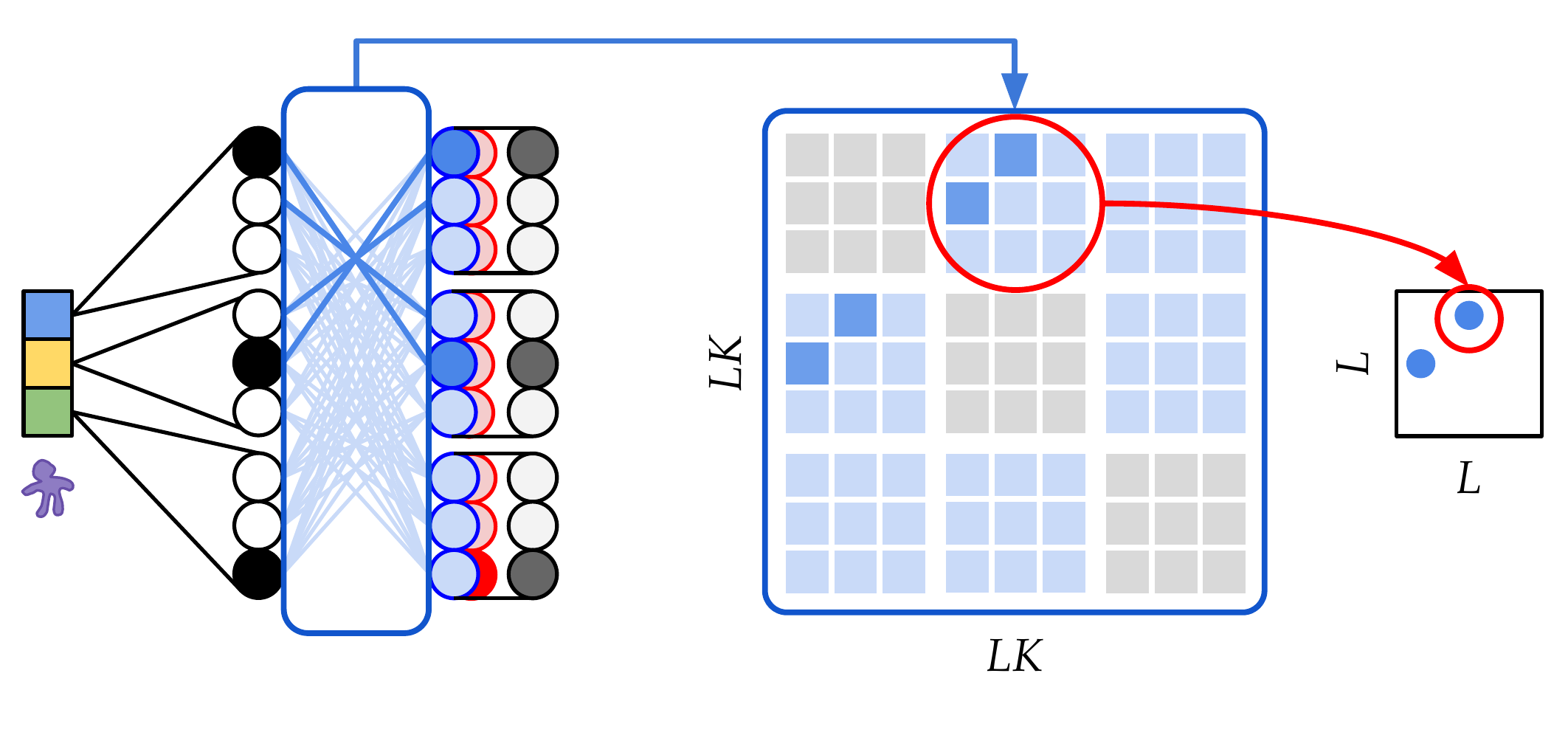}
  \caption{From the Markov Random Field model with pseudo-likelihood one can infer the matrix $\mathbf{W}\in \mathbb{R}^{LK\times LK}$ and take L2 norm to get a contact map shown on the right.}
  \label{fig:3}
\end{figure}

In the case of MRFs (Markov Random Fields) , the two-body term is used for graph or contact map inference (Fig. \ref{fig:3}). For the mean-field approximation of MRFs \cite{morcos2011direct} or MG models \cite{baldassi2014fast}, the precision (or inverse covariance) matrix is used to infer "direct causation" \cite{markowetz2007inferring}. Though the application is similar, it is not immediately obvious why maximizing the pseudo-likelihood of a regularized MRF results in a more accurate pairwise term \cite{balakrishnan2011learning, kamisetty2013assessing}, compared to estimating this term by taking the inverse of the shrunken covariance matrix.
Under our unifying framework, the major difference between these models is their loss functions. Beyond that,  different regularization techniques are used to promote sparsity in MSE and CCE models. For MSE, these include the addition of a pseudo-count, small constant along the diagonal \cite{morcos2011direct, baldassi2014fast, rawi2016couscous} and/or L1 penalty during graphical lasso optimization \cite{friedman2008sparse, jones2011psicov}. For CCE, these include L1, group L1 and L2 penalty, where L2 regularization was found to work best for contact prediction \cite{kamisetty2013assessing, ekeberg2013improved}. 
\begin{figure}
  \centering
  \includegraphics[height=3.3cm]{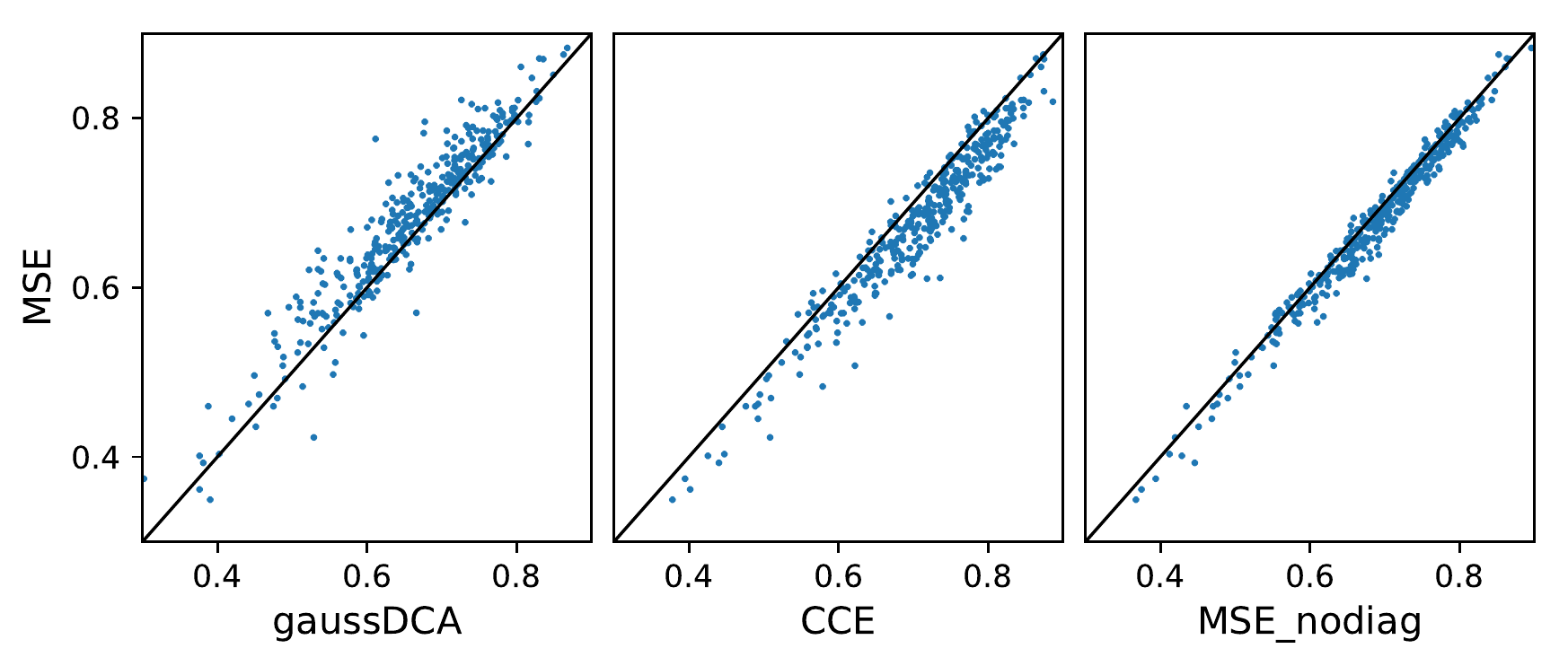}
  \caption{MSE is more accurate than GaussDCA \cite{baldassi2014fast} for protein contact prediction, but worse than CCE. Each point corresponds to a protein. The axes indicate the accuracy of each method, defined as the average precision of the top L ranked contacts. L is the length of the protein, the contact is defined using CONFIND \cite{holland2018contact}, between every pair of positions that are greater than or equal to sequence separation of 6.}
  \label{fig:4}
\end{figure}
To test if L2 regularization also works for MSE, we add a small constant $4.5/N^{1/2}$ along the diagonal before inversion. Remarkably, MSE outperforms GaussDCA \cite{baldassi2014fast}, which adds a pseudo-count to the entire covariance matrix. By zeroing the diagonal of the pairwise matrix (see Eq. \ref{eq:13}), we see a small but consistent improvement. Coincidentally, we find this to be equivalent to computing the partial correlation coefficient. Though the results are still not as good as CCE, MSE with equivalent regularization provides a closer approximation to CCE (see Fig. \ref{fig:4}).

The test was performed on a diverse set of proteins with at least 1K effective sequences from \cite{anishchenko2017origins}. The redundancy was reduced by selecting one random member per connected component, where the edges of the graph were defined using a 1E-10 e-value threshold, computed by HMM-HMM alignment \cite{hildebrand2009fast}.  To keep the data-set consistent, each alignment was sub-sampled to 1K.  Before ranking, all methods were corrected for entropy using Average Product Correction \cite{dunn2007mutual}.
\section{Conclusion}
We showed that widely used models for multivariate distributions in biological sequences can all be expressed as a single fully-connected layer, where the weights and bias of the dense layer captures the co-evolution and conservation, respectively. The differences in the models comes down to the loss function used, where inverse-covariance methods are effectively minimizing the mean-squared error (more appropriate for continuous data) and markov-random-field methods are minimizing the categorical cross entropy (more appropriate for categorical data). Since maximizing the pseudo-likelihood for MRF is identical to minimizing the categorical cross entropy (loss function often used for encoding/decoding biological sequences) this further reduces the difference between AE and MRF to a simple addition of hidden layer(s) between the input and output.

Though the result is simple, it helps unite and bridge the gap  between these widely used models across multiple fields. This framework should allow researchers to easily extend and incorporate features used in one set of models to others. We show a simple example of this by incorporating regularization often used in MRFs to MG models.

\paragraph{Acknowledgements}
MN is supported by the James S. McDonnell Foundation, Schmidt Futures, and Israel Council for Higher Education. MN and SO are supported by the John Harvard Distinguished Science Fellows Program within the FAS Division of Science of Harvard University. Research reported in this publication was supported by Office of the Director of the National Institutes of Health under award number DP5OD026389. The content is solely the responsibility of the authors and does not necessarily represent the official views of the National Institutes of Health.


\bibliography{sergey}
\bibliographystyle{icml2019}
\blfootnote{Code is available at \href{https://github.com/sokrypton/seqmodels}{https://github.com/sokrypton/seqmodels}}

\end{document}